\documentclass[twocolumn,english,aps,prb,twocolumn,superscriptaddress,bibnotes,amsmath,amssymb,floatfix]{revtex4-1}
\usepackage[colorlinks=true,citecolor=blue,linkcolor=magenta]{hyperref}

\usepackage[flushleft]{threeparttable}
\usepackage[table]{xcolor}

\usepackage[utf8]{inputenc}
\usepackage[english]{babel}
\usepackage{amsmath,amsfonts,amssymb}
\usepackage[T1]{fontenc}
\usepackage{xurl}
\usepackage{soul}
\usepackage{changes}

\usepackage{amsmath}
\usepackage{amsfonts}
\usepackage{amssymb}
\usepackage[version=3]{mhchem}
\usepackage{stmaryrd}
\usepackage{epstopdf}
\usepackage{graphicx}
\graphicspath{{./Figures/}}

\definecolor{mygray}{gray}{0.95}

\begin{document}

\title{Photo-induced cascaded harmonic and comb generation in \\ silicon nitride microresonators}

\author{Jianqi Hu}
\thanks{These authors contributed equally to the work.}
\affiliation{{\'E}cole Polytechnique F{\'e}d{\'e}rale de Lausanne, Photonic Systems Laboratory (PHOSL), STI-IEM, Station 11, Lausanne CH-1015, Switzerland.}

\author{Edgars Nitiss}
\thanks{These authors contributed equally to the work.}
\affiliation{{\'E}cole Polytechnique F{\'e}d{\'e}rale de Lausanne, Photonic Systems Laboratory (PHOSL), STI-IEM, Station 11, Lausanne CH-1015, Switzerland.}

\author{Jijun He}
\affiliation{{\'E}cole Polytechnique F{\'e}d{\'e}rale de Lausanne, Laboratory of Photonics and Quantum Measurements (LPQM), SB-IPHYS, Station 3, Lausanne CH-1015, Switzerland.}

\author{Junqiu Liu}
\affiliation{{\'E}cole Polytechnique F{\'e}d{\'e}rale de Lausanne, Laboratory of Photonics and Quantum Measurements (LPQM), SB-IPHYS, Station 3, Lausanne CH-1015, Switzerland.}

\author{Ozan Yakar}
\affiliation{{\'E}cole Polytechnique F{\'e}d{\'e}rale de Lausanne, Photonic Systems Laboratory (PHOSL), STI-IEM, Station 11, Lausanne CH-1015, Switzerland.}

\author{Wenle Weng}
\affiliation{{\'E}cole Polytechnique F{\'e}d{\'e}rale de Lausanne, Laboratory of Photonics and Quantum Measurements (LPQM), SB-IPHYS, Station 3, Lausanne CH-1015, Switzerland.}

\author{Tobias J. Kippenberg}
\affiliation{{\'E}cole Polytechnique F{\'e}d{\'e}rale de Lausanne, Laboratory of Photonics and Quantum Measurements (LPQM), SB-IPHYS, Station 3, Lausanne CH-1015, Switzerland.}

\author{Camille-Sophie Br\`es}
\email[]{camille.bres@epfl.ch}
\affiliation{{\'E}cole Polytechnique F{\'e}d{\'e}rale de Lausanne, Photonic Systems Laboratory (PHOSL), STI-IEM, Station 11, Lausanne CH-1015, Switzerland.}

\maketitle

\noindent\textbf{\noindent
Silicon nitride (Si$_3$N$_4$) is an ever-maturing integrated platform for nonlinear optics. 
Yet, due to the absence of second-order ($\chi^{(2)}$) nonlinearity, Si$_3$N$_4$ is mostly considered for third-order ($\chi^{(3)}$) nonlinear interactions.
Recently, this limitation was overcome by optical poling in both Si$_3$N$_4$ waveguides and microresonators via the photogalvanic effect, resulting in the inscription of quasi-phase-matched $\chi^{(2)}$ gratings. 
Here, we report cascaded nonlinear effects in a normal dispersion Si$_3$N$_4$ microresonator with combined $\chi^{(2)}$ and $\chi^{(3)}$ nonlinearities. 
We demonstrate that the photo-induced $\chi^{(2)}$ grating also provides phase-matching for the sum-frequency generation process, enabling the initiation and successive switching of primary combs at pump wavelength. Additionally, the doubly resonant pump and second-harmonic fields allow for cascaded third-harmonic generation, where a secondary optically written $\chi^{(2)}$ grating is identified. Finally, we reach a low-noise, broadband microcomb state evolved from the sum-frequency coupled primary comb. These results expand the scope of cascaded effects in $\chi^{(2)}$ and $\chi^{(3)}$ microresonators. 
}

\section*{Introduction} 

\noindent{Optical} microresonators made of low-loss and highly nonlinear materials have emerged as prominent systems for the observation and the understanding of the rich landscape of nonlinear physics. Typical nonlinear interactions in microresonators encompass second-order ($\chi^{(2)}$)\cite{breunig2016three} and third-order ($\chi^{(3)}$)\cite{moss2013new} parametric processes, as well as photon-phonon processes such as Raman\cite{spillane2002ultralow} and Brillouin\cite{eggleton2019brillouin} scatterings. Specifically, $\chi^{(3)}$ nonlinearity is ubiquitous in all material platforms and thus mostly exploited, enabling third-harmonic (TH) generation\cite{surya2018efficient,li2020third} or Kerr frequency comb generation\cite{gaeta2019photonic,pasquazi2018micro}. The past decades have witnessed extensive microcomb studies, evolving rapidly from the primary combs\cite{kippenberg2004kerr} to broadband, mode-locked states where dissipative Kerr solitons\cite{herr2014temporal} or dark pulses\cite{xue2015mode} are formed. This has subsequently found numerous applications thanks to the mutually coherent and massively parallel microcomb lines\cite{kippenberg2018dissipative}.

Unlike the widely accessible Kerr effect in microresonators, $\chi^{(2)}$ nonlinearity is only intrinsic to non-centrosymmetric media. Notably, $\chi^{(2)}$ response is essential for electro-optic effect\cite{zhu2021integrated} and underpins various three-wave mixing parametric processes, for example second-harmonic (SH) generation\cite{lin2019broadband,chen2019ultra,lu2020toward,bruch201817,kuo2014second,chang2019strong, logan2018400}. 
Efficient SH has been generated in $\chi^{(2)}$ microresonators in lithium niobate\cite{lin2019broadband,chen2019ultra,lu2020toward} and III-V materials\cite{bruch201817,kuo2014second,chang2019strong, logan2018400}. Beyond SH generation, cascaded $\chi^{(2)}$ processes resulting in comb formation have been observed\cite{szabados2020frequency,hendry2020experimental}, as $\chi^{(2)}$ cavities also display effective $\chi^{(3)}$ nonlinearity\cite{ricciardi2015frequency,mosca2016direct,leo2016walk}. Additionally, parametrically driving $\chi^{(2)}$ microresonators could allow for the onset of optical parametric oscillation\cite{beckmann2011highly,bruch2019chip,szabados2020frequency_apl,lu2021ultralow}. This has further led to the recent observation of quadratic solitons in $\chi^{(2)}$ microresonators\cite{bruch2021pockels} akin to localized structures in $\chi^{(3)}$ microresonators\cite{kippenberg2018dissipative}.

However, with the exception of silicon carbide\cite{lukin20204h,wang2021high}, conventional silicon photonic materials (e.g. silicon, silica and silicon nitride) lack intrinsic $\chi^{(2)}$ response. In order to bring $\chi^{(2)}$ into these complementary metal–oxide–semiconductor (CMOS)-compatible materials, diverse approaches have been pursued\cite{timurdogan2017electric,singh2020broadband,castellan2019origin,billat2017large,porcel2017photo,hickstein2019self,nitiss2019formation,nitiss2020broadband,zhang2019symmetry}. Paradigm demonstrations include symmetry-breaking in silicon waveguides induced by applied electric fields\cite{timurdogan2017electric,singh2020broadband} or via charged defects\cite{castellan2019origin} and all-optical poling of silicon nitride (Si$_3$N$_4$) waveguides based on the photogalvanic effect\cite{billat2017large,porcel2017photo,hickstein2019self,nitiss2019formation,nitiss2020broadband}. With regard to resonator structures, effective $\chi^{(2)}$ can also be optically induced in Si$_3$N$_4$ microresonators manifesting in efficient SH generation\cite{levy2011harmonic,lu2021efficient,nitiss2022optically}. 

SH generation in microresonators without periodically poled\cite{chen2019ultra,lu2020toward} or crystallographic\cite{kuo2014second} quasi-phase-matching (QPM) structures, had long been achieved by intermodal phase-matching between involved fundamental-harmonic (FH, i.e., pump) and SH resonances. 
Opposed to this convention, we have recently observed in Si$_3$N$_4$ that periodic space-charges are self-organized along the circumferences of microresonators when the pump and its SH are doubly resonant\cite{nitiss2022optically}. 
The alternating electric field together with the $\chi^{(3)}$ nonlinearity of Si$_3$N$_4$ results in an effective $\chi^{(2)}$ grating\cite{anderson1991model,dianov1995photoinduced,margulis1995imaging}, with a periodicity following the interference of FH and SH, and thus automatically compensating their phase-mismatch. Such fully adaptive QPM greatly eases the efforts for meeting phase-matching condition, posing Si$_3$N$_4$ microresonators as ideal testbeds to investigate various $\chi^{(2)}$ nonlinear effects beyond SH generation. Meanwhile, Si$_3$N$_4$ is indisputably one of the most popular platforms for Kerr microcomb research\cite{ferdous2011spectral,xue2015mode,brasch2016photonic,stern2018battery,liu2020photonic,xiang2021laser,helgason2022power}. The combined $\chi^{(2)}$ and $\chi^{(3)}$ combs have also been observed in Si$_3$N$_4$ microresonators\cite{miller2014chip,xue2017second}, albeit the presence of $\chi^{(2)}$ being attributed to intermodal phase-matching. 

In this paper, we demonstrate versatile $\chi^{(2)}$ and $\chi^{(3)}$ nonlinear effects in a high quality factor (\textit{Q}), normal dispersion Si$_3$N$_4$ microresonator. A tunable continuous-wave (CW) laser is injected into the fundamental pump resonances of the microresonator. By varying the pump detuning, the doubly resonant condition of the pump mode and one of the several SH modes is easily matched, thanks to the small free spectral range (FSR $\sim25~{\rm GHz}$) and multimode nature of the microresonator used. This gives rise to the photo-induced $\chi^{(2)}$ grating inside the microresonator enabling QPM for SH generation\cite{nitiss2022optically}. Following the FH-SH grating inscription, we here show that the combined $\chi^{(2)}$ and $\chi^{(3)}$ nonlinearities can lead to the cascaded nonlinear processes like the generation of TH and frequency combs. In our system, the TH generation is also a $\chi^{(2)}$ process resulting from the photogalvanic effect, but in this case based on the coherent interference among the FH, the generated SH and the seed TH waves. We confirm this by using two-photon microscopy (TPM) imaging of the poled microresonator\cite{hickstein2019self,nitiss2019formation,nitiss2022optically}, where a secondary $\chi^{(2)}$ grating responsible for the QPM of the underlying FH-SH-TH process is clearly identified. Additionally, we observe the generation of a primary comb in the FH band together with sum-frequency (SF) generation in the SH band. The first-order comb sideband and the SF component are efficiently coupled through the SF generation process and can be successively switched while tuning the pump further into the resonance. Here the phase-matching of the SF generation process is the same as the SH generation, thereby guaranteed by the already inscribed FH-SH grating. Besides, the SF coupled primary comb also facilitates the excitation of localized states in the normal dispersion regime\cite{parra2016origin}. As a consequence, a low-noise and broadband microcomb featuring typical spectral envelope of a dark pulse is deterministically reached in the experiment. Our work reports novel cascaded nonlinear phenomena in an optically poled Si$_3$N$_4$ microresonator, and is potentially applicable for $f$-$2f$ or $2f$-$3f$ comb self-referencing\cite{jung2014green,he2019self}. 

\begin{figure*}[hbt!]
  \centering{
  \includegraphics[width = 1\linewidth]{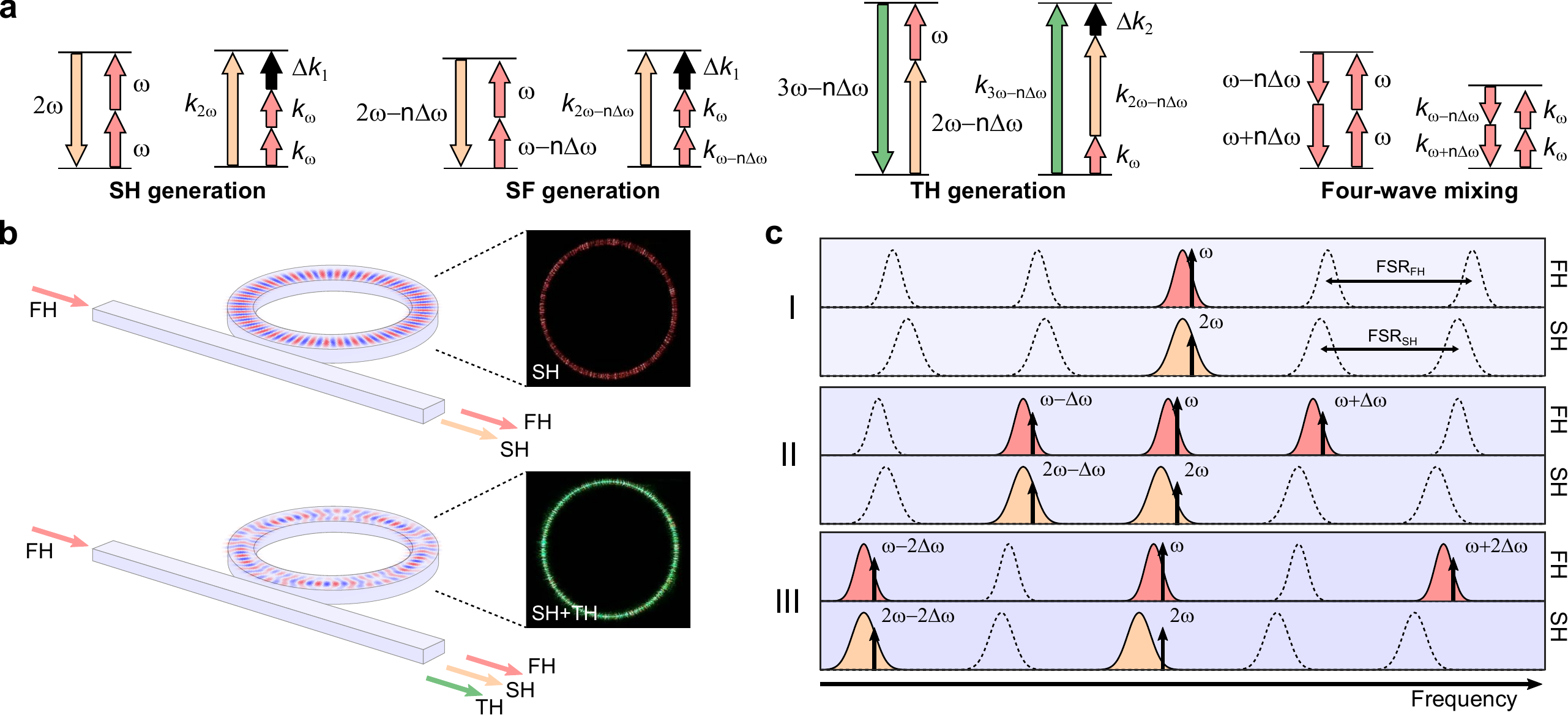}
  } 
    \caption{\noindent\textbf{Photogalvanic effect enabled processes in Si$_3$N$_4$ microresonators.} \textbf{a}. Illustration of momentum and energy conservation of the available processes inside the device: second-harmonic (SH) generation, sum-frequency (SF) generation, third-harmonic (TH) generation, and four-wave mixing (FWM). SH, SF, and TH generation processes are present due to quasi-phase-matching (QPM) of participating waves enabled by the photogalvanic effect. FWM process is initiated with the assistance of SF coupling.  
    \textbf{b} Artistic representation of $\chi^{(2)}$ QPM grating structures optically inscribed in the Si$_3$N$_4$ microresonator. 
    The top part illustrates the FH-SH grating inscription when SH is generated, while the bottom part illustrates the inscription of superimposed FH-SH and FH-SH-TH gratings when SH and TH are both generated. 
    Optical microscope images of the scattered light from the microresonator are shown for both cases. 
    \textbf{c} Relative resonance distribution at fundamental-harmonic (FH) and SH bands in SF coupled FWM process.
    By tuning the pump frequency $\omega$ into the resonance (decreasing $\omega$), the SH resonances red-shift faster than the FH resonances.
    During the detuning the doubly resonant condition for pump $\omega$ and its SH 2$\omega$ is met, leading to $\chi^{(2)}$ grating inscription and SH generation (I).
    When the pump is further detuned, the frequency-matching of the SF process among modes $\omega$, $\omega-\Delta\omega$ and $2\omega-\Delta\omega$ is achieved, while its phase-matching is guaranteed by the already inscribed $\chi^{(2)}$ grating. 
    The SF coupled FWM occurs manifested by the simultaneous generation of sideband modes $\omega\pm\Delta\omega$ at FH and SF mode $2\omega-\Delta\omega$ at SH (II). 
    The coupled SF mode can be successively downshifted, and the free spectral range (FSR) of the FH comb is varied accordingly (III). 
  }
 \label{Figure1}
\end{figure*} 

\section*{Results} 
\noindent \textbf{$\chi^{(2)}$ and $\chi^{(3)}$ nonlinear processes in Si$_3$N$_4$ microresonators.} In this study, we employ a large radius, normal dispersion Si$_3$N$_4$ microresonator (see Methods for device details) with no significant mode crossings in the pump spectral region (see Supplementary Note 1). Kerr comb generation in such microresonators is considered difficult with typical CW driving\cite{godey2014stability}, as combs in normal dispersion regime are generally initiated by leveraging perturbations from spatial mode coupling\cite{liu2014investigation,xue2015normal} or self-injection locking\cite{jin2021hertz,lihachev2021platicon}. However, in this work, we show that multiple nonlinear processes can be enabled by the photogalvanic effect, resulting in harmonic generation as well as comb generation. Figure \ref{Figure1}a illustrates the energy and momentum conservation diagrams of these processes. In the experiment, a CW pump at angular frequency $\omega$ is slowly scanned by decreasing its frequency to stably access a pump resonance. While sweeping, if a certain SH mode resonance matches the frequency $2\omega$, the microresonator becomes doubly resonant and the photogalvanic effect could occur\cite{lu2021efficient,nitiss2022optically}. The phase-matching is automatically realized via a self-organized $\chi^{(2)}$ grating, inscribed following the interference of resonating FH and initially very weak SH. We can rewrite the phase mismatch as ${\Delta k_1 = k_{2\omega} - 2k_{\omega} = (m_{2\omega} - 2m_{\omega})/R }$\cite{nitiss2022optically} with $m_{\omega}$ and $m_{2\omega}$ the azimuthal mode numbers of FH and SH resonances, respectively, and $R$ the radius of the microresonator. The corresponding QPM condition is given by $|\Delta k_1| = 2\pi/\Lambda_{1}$ with $\Lambda_{1}$ the grating period. Hence, the grating period length is derived as $\Lambda_{1}=2\pi R/|m_{2\omega} - 2m_{\omega}|$, and the total number of grating periods per perimeter writes $N_1 = |m_{2\omega} - 2m_{\omega}|$. 
As depicted by the artistic representation in Figure \ref{Figure1}b (top), the $\chi^{(2)}$ QPM grating in turn drastically amplifies the process, leading to efficient SH generation.

After the inscription of the FH-SH grating for SH generation, a series of cascaded nonlinear processes follow when the pump frequency is further scanned within the resonance. These include SF generation, comb generation, and TH generation as detailed in Fig. \ref{Figure1}a. The SF and primary comb generation is explained in Fig. \ref{Figure1}c, where the relative distributions of FH and SH resonances at different pump detunings are shown. 
Initially, the doubly resonant condition is fulfilled so that frequency $2\omega$ is generated (region I). Here the FSRs of different SH modes are smaller than the pump resonance FSR (see Supplementary Note 1), and SH resonances red-shift faster than the pump resonance while detuning\cite{nitiss2022optically}. As a consequence, the frequency-matching among the FH mode at $\omega$, FH sideband mode at $\omega-\Delta\omega$, and SF mode at $2\omega-\Delta\omega$, with $\Delta\omega$ approximately the FSR at pump, can be realized (region II). It is important to note that such a three-wave process is also phase matched, given that the azimuthal mode number mismatch among three waves is the same as the number of inscribed grating periods in SH generation, i.e. $|(m_{2\omega}-1)-m_\omega-(m_\omega-1)|=N_1$. The nonlinear coupling via the SF process modifies locally the phase of this particular sideband mode, enabling degenerate four-wave mixing (FWM) even in the normal dispersion regime and simultaneous SF generation\cite{xue2017second}. If further pump detuning is possible, the SF coupled mode at FH will be successively downshifted to $\omega-n\Delta\omega$, leading to primary comb generation with comb FSR of $n\Delta\omega$ and SF generation at a frequency of $2\omega-n\Delta\omega$ (region III, $n=2$ is shown). Flexible primary comb FSR switching can thereby be obtained in a single pump resonance via the dynamic coupling. This is in contrast to the conventional spatial mode induced coupling\cite{liu2014investigation,xue2015normal}, where the crossing position is generally fixed so that the comb FSR is varied by pumping different resonances. We also theoretically analyze the modulational instability (MI) of the process based on coupled Lugiato–Lefever equations\cite{leo2016walk,leo2016frequency,xue2017second} (see Supplementary Note 2). Net gain is indeed observed in simulation with the peak position increasing with the pump detuning, confirming the phenomenon depicted in Fig. \ref{Figure1}c. 

The generation of TH also takes place in the microresonator given the presence of the SH field. Unlike the typical approaches based on the intermodal phase-matched $\chi^{(3)}$ process\cite{surya2018efficient,wang2016frequency}, here the TH is induced by the photogalvanic effect based on the interference among FH, SH, and TH waves. Indeed, ref. \cite{yakar2022coherent} has highlighted that three arbitrary coherent waves that are frequency matched could yield a periodic electric field to compensate for their unmatched momenta. Hence, as shown in Fig. \ref{Figure1}a, the SH and TH waves here can also be offsetted at frequencies of $2\omega- n\Delta\omega$ and $3\omega- n\Delta\omega$, respectively. This offset depends on the resonant condition of the waves, as similar to SH generation, and the process now requires triply resonant in order to be initiated. Under such a circumstance, the interference of the three waves results in a self-organized FH-SH-TH grating due to the photogalvanic effect. Once again, the number of grating periods will exactly compensate the azimuthal mode number mismatch among three waves, in this case given by $N_2 = |(m_{3\omega}-n)-(m_{2\omega}-n)-m_{\omega}| = |m_{3\omega}-m_{2\omega}-m_{\omega}|$, or equivalently the phase mismatch $\Delta k_2 =(m_{3\omega} - m_{\omega} - m_{2\omega})/R$. Therefore, there should be two superimposed periodic electric fields written in the microresonator that are respectively phase matched for the FH-SH and FH-SH-TH processes, as illustrated by the artistic representation in Fig. \ref{Figure1}b (bottom).

 \begin{figure*}[htp]
  \centering{
  \includegraphics[width = 0.88\linewidth]{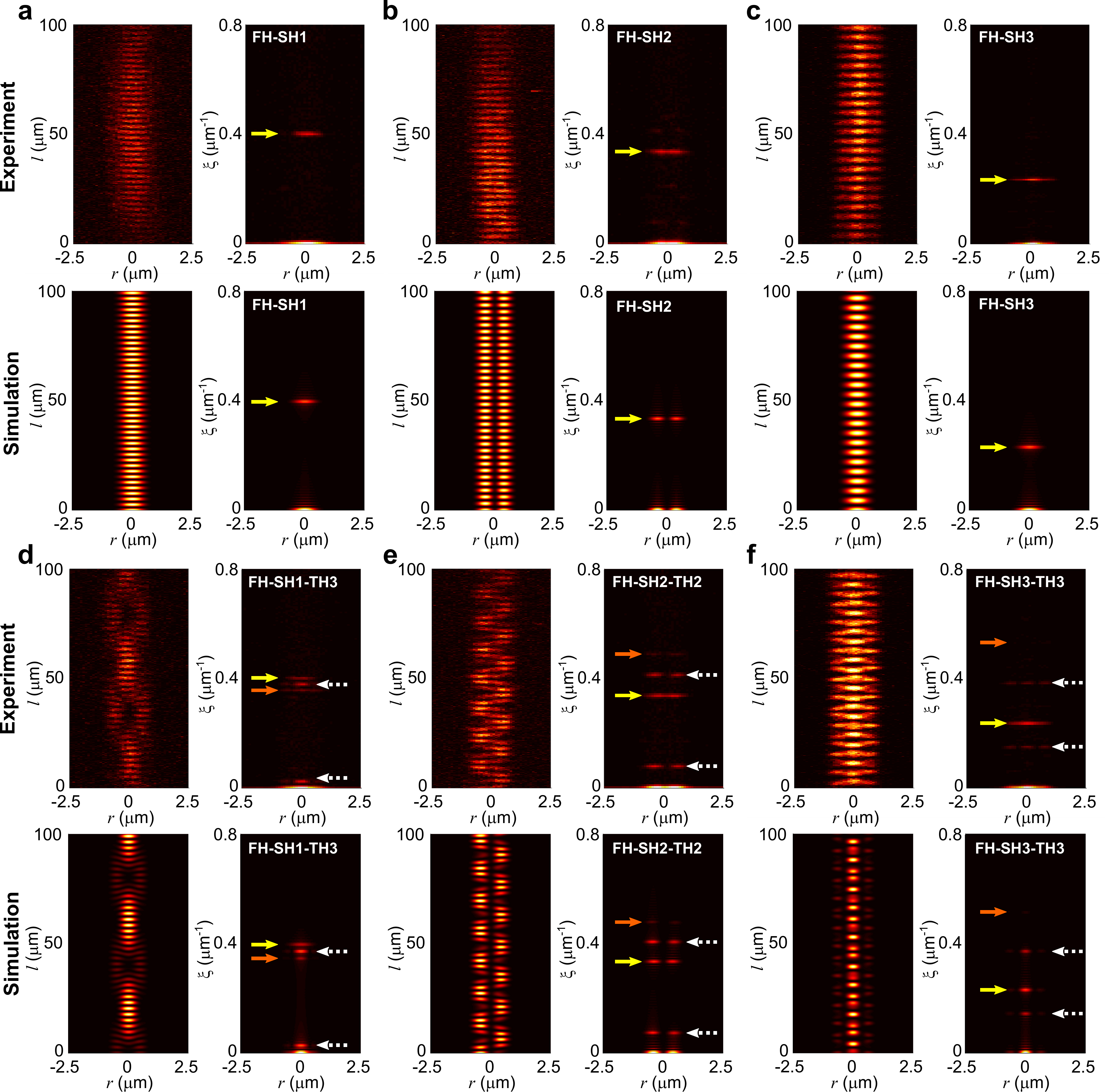}
  } \caption{\noindent\textbf{Imaging and identification of inscribed $\chi^{(2)}$ gratings in the Si$_3$N$_4$ microresonator.} 
  \textbf{a-c} FH-SH grating structures (plotted along the circumference of the microresonator) and their spatial frequency graphs obtained by the Fourier transform. Top: experiment; Bottom: simulation. 
  The gratings in \textbf{a-c} are inscribed by slightly tuning the pump into the resonances near $1544.40~{\rm nm}$, $1547.35~{\rm nm}$ and $1552.34~{\rm nm}$, respectively. Based on the retrieved spatial frequencies (yellow arrows), the participating SH modes in the FH-SH photogalvanic process are identified as SH1, SH2, SH3, respectively. 
  \textbf{d-f} Superimposed FH-SH and FH-SH-TH grating structures and their spatial frequency graphs. Top: experiment; Bottom: simulation. The gratings in \textbf{d-f} are inscribed by further tuning the pump into the same resonances as in \textbf{a-c}, respectively. Multiple Fourier components are observed corresponding to the spatial frequencies of the FH-SH gratings (yellow arrows, same as in \textbf{a-c}), the FH-SH-TH gratings (orange arrows), and the interference of the two gratings (dashed white arrows). In \textbf{d-f}, the TH modes involved in the FH-SH-TH photogalvanic process are identified as TH3, TH2, and TH3, respectively.  
 }
  \label{Figure2}
\end{figure*} 

\noindent\textbf{$\chi^{(2)}$ grating imaging and analysis.} The $\chi^{(2)}$ grating in the microresonator is inscribed by coupling a CW pump into resonances at C-band (see Supplementary Note 3). After such optical poling, we use TPM imaging technique to characterize the grating structure\cite{hickstein2019self,nitiss2019formation,nitiss2022optically}. The detail of TPM measurement and the following image processing are described in the Methods section. In essence, the approach allows for the extraction of $\chi^{(2)}$ grating shape along the circumference of microresonator as well as their spatial frequency through spatially-resolved Fourier analysis. 
This enables precise identification of the grating periods, which is essential for determining the modes interacting during the photogalvanic process.

Figure \ref{Figure2} showcases several processed $\chi^{(2)}$ grating images after the coordinate transformation (see Methods). All these diverse grating images are attained from the same microresonator, but are written at three different pump resonances with different detuning conditions. Figures \ref{Figure2}a-c correspond to the $\chi^{(2)}$ gratings inscribed when the pump is slightly tuned into resonances near $1544.40~{\rm nm}$, $1547.35~{\rm nm}$ and $1552.34~{\rm nm}$, respectively. Only SH is generated (see Fig. \ref{Figure1}b photo of the microresonator glowing red) in these cases, and no cascaded nonlinear effect is observed.
The grating structure in the microresonator is given by the nonlinear interference of the involved FH and SH waves, i.e. $\chi^{(2)}(\phi) \sim  \operatorname{Re}\{E_{2\omega}(E_{\omega}^*)^2 e^{i\Delta k_1 R \phi}\} $, where $E_{\omega}$ and $E_{2\omega}$ are the optical fields at FH and SH, respectively. $ \operatorname{Re}\{ \}$ is the real part of a complex number and  $^*$ stands for the complex conjugate. In TPM images the actual measured SH intensity is proportional to $(\chi^{(2)}(\phi))^2$. The mode profiles and effective refractive indices of FH and several lower-order SH modes (SH1 to SH3) are extracted from COMSOL Multiphysics simulation (see Supplementary Note 1). Based on these, we can simulate the corresponding grating patterns\cite{nitiss2022optically} and identify their participating SH modes in Figs. \ref{Figure2}a-c to be SH1, SH2, and SH3, respectively. The Fourier transform additionally confirms that there is only one dominant spatial frequency for each case ($2|\Delta k_1|$, indicated by the yellow arrow).

When the pump is further tuned into the same resonances near $1544.40~{\rm nm}$, $1547.35~{\rm nm}$ and $1552.34~{\rm nm}$, cascaded nonlinear effects including TH generation are observed while SH generation is still maintained (see Fig. \ref{Figure1}b photo of the microresonator glowing green). 
To verify that the TH results from the photogalvanic induced $\chi^{(2)}$ nonlinearity instead of intermodal $\chi^{(3)}$ process, we again image the poled microresonator generating TH using the TPM. Complex $\chi^{(2)}$ grating patterns of "chain", "zipper", and "necklace" shapes are captured and shown in Figs. \ref{Figure2}d-f, respectively. From the Fourier transform we can decompose the superimposed grating structures into their Fourier components, where multiple spatial frequencies are now identified, in contrast to the single spatial frequency observed in Figs. \ref{Figure2}a-c. 
Considering that the $\chi^{(2)}$ grating is composed of two parts separately phase-matched for FH-SH and FH-SH-TH generation processes, the $\chi^{(2)}$ response can be written as $\chi^{(2)}(\phi) \sim \operatorname{Re}\{E_{2\omega}(E_{\omega}^*)^2 e^{i\Delta k_1 R \phi} + a_0 E_{3\omega}E_{\omega}^* E_{2\omega}^* e^{i\Delta k_2 R \phi - i\phi_0}\}$, where $E_{3\omega}$ is the resonant TH field, $a_0$ and $\phi_0$ are the relative amplitude and phase between the two gratings, respectively. We have assumed that the participating SH (TH) field here in the FH-SH-TH photogalvanic process is at frequency $2\omega (3\omega)$ as its mode profile and effective index barely change with a small offset $n$. We clearly see that the $(\chi^{(2)}(\phi))^2$ responses measured which possess four nonzero spatial frequencies, i.e. $2|\Delta k_1|$, $2|\Delta k_2|$, $|\Delta k_1 + \Delta k_2|$ and $|\Delta k_1 -\Delta k_2|$. 
We note that one of the spatial frequencies in Figs. \ref{Figure2}d-f, indicated again by the yellow arrow, is identical to the frequency extracted from Figs. \ref{Figure2}a-c. This implies that the participating SH mode remains the same within each pump resonance while varying the pump detuning. 
Afterwards, we identify in Figs. \ref{Figure2}d-f the spatial frequency of the secondary grating ($2|\Delta k_2|$, orange arrow) and the interference between the two gratings ($|\Delta k_1 + \Delta k_2|$, $|\Delta k_1 -\Delta k_2|$, dashed white arrows) according to their frequency relations. 
In case the secondary grating is weak, its spatial frequency can still be easily retrieved based on the interference components. 
From the extracted secondary frequency we infer the involved TH mode in the photogalvanic process, and then reconstruct the grating shape using simulated mode profiles (see Supplementary Note 1). By considering the involved TH modes to be TH3, TH2, and TH3 in Figs. \ref{Figure2}d-f, respectively, the simulated grating structures are reproduced very well with measurement results both in terms of grating shapes and periods.

\begin{figure*}[hbt!]
  \centering{
  \includegraphics[width = 1.0\linewidth]{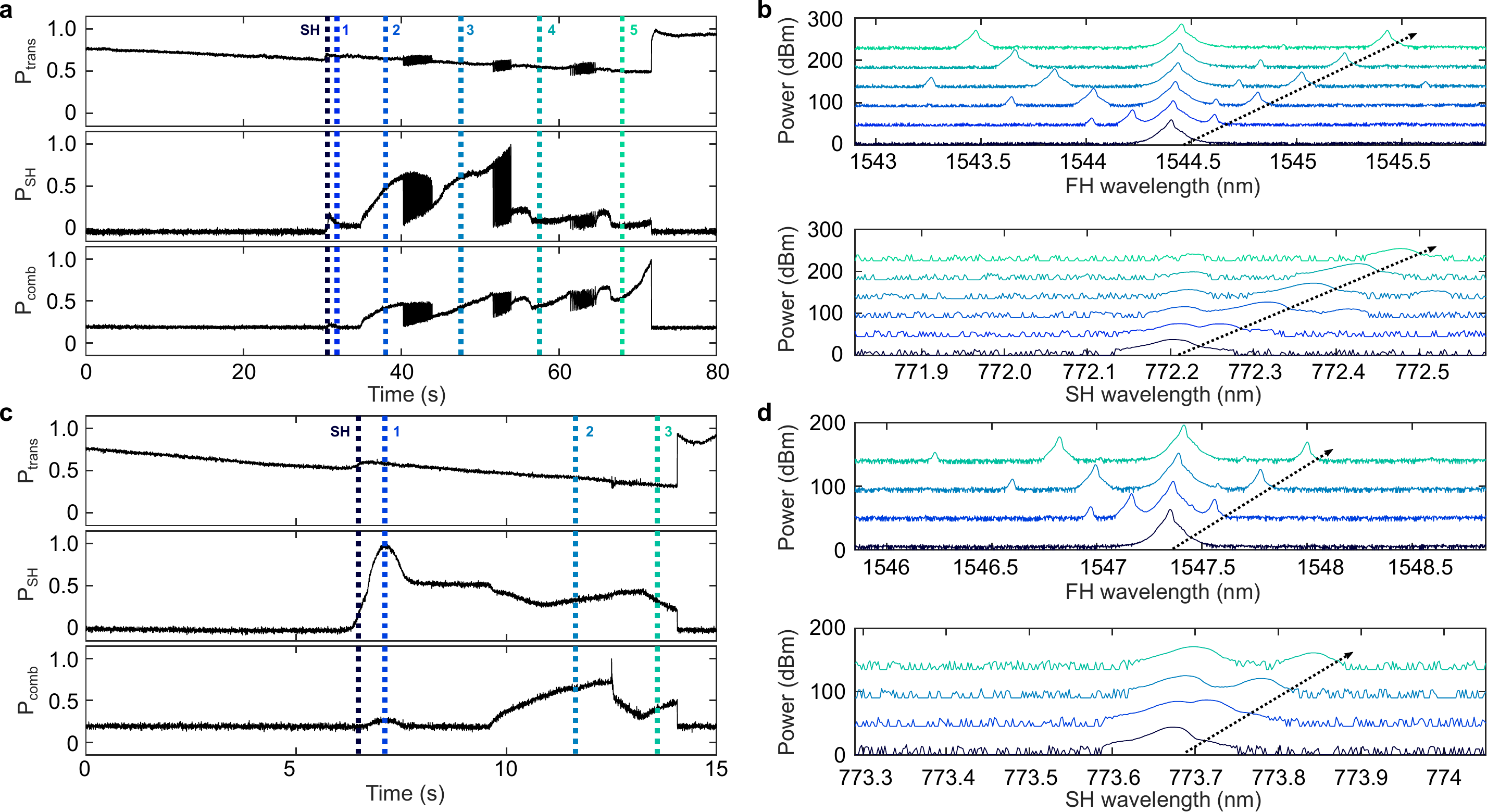}
  } 
    \caption{\noindent\textbf{Consecutive switching of primary comb FSR and sum-frequency mode.}
    \textbf{a} and \textbf{c} Evolution of pump transmission ($P_{\rm trans}$), generated SH power ($P_{\rm SH}$) and comb power ($P_{\rm comb}$) during the pump wavelength sweeps across the resonance at $1544.40~{\rm nm}$ and $1547.35~{\rm nm}$, respectively.   
    \textbf{b} and \textbf{d} The measured FH (top) and SH (bottom) spectra at the wavelength regions indicated by the vertical colored lines in \textbf{a} and \textbf{c}. 
    After the inscription of the $\chi^{(2)}$ grating (SH generation), a primary comb is initiated with the simultaneous generation of the SF component. 
    The FSR of the primary comb and the generated SF are then successively switched. The dashed lines in \textbf{b} and \textbf{d} denote the shifting of the coupled sideband modes at FH and SH bands. 
  }
 \label{Figure3}
\end{figure*} 

\noindent \textbf{Switchable primary comb and sum-frequency generation.}
The photo-induced $\chi^{(2)}$ gratings also allow for the primary comb and SF generation in the normal dispersion microresonator. To characterize these, we use an experimental setup that measures the transmission curves as well as the generated optical spectra at both pump and SH wavelengths (see Supplementary Note 3). When the pump frequency is slowly scanned across the $1544.40~{\rm nm}$ resonance, the transmission curves of pump, SH, and comb power ($P_{\rm trans}$, $P_{\rm SH}$ and $P_{\rm comb}$, respectively) are recorded and plotted in Fig. \ref{Figure3}a. Figure \ref{Figure3}b shows the measured FH and SH spectra at different pump detunings corresponding to the vertical dashed lines in Fig. \ref{Figure3}a. These spectral plots are stacked and shifted in the y-axis by $45~{\rm dB}$ for clarity. From the transmission curves, we see that the primary comb generation at FH is initiated after SH generation. 
Further laser tuning into the resonance results in FSR switching of the primary combs from 1 FSR to 5 FSR, in steps of 1 FSR (guided by the dashed arrow), before the pump exists its resonance. In the meantime, we can clearly observe the generation of the SF mode at the SH band, whose mode index is also downshifted by 1 FSR to 5 FSR away from the SH mode (guided by the dashed arrow). Additionally, we notice that the generated primary comb spectra are asymmetric, where the first-order lower sideband $\omega - n\Delta\omega$ is always weaker than the upper sideband $\omega + n\Delta\omega$ for a $n$ FSR comb ($n =1,...,5$). This actually implies that the sideband mode $\omega - n\Delta\omega$ is phase-matched\cite{huang2017globally}, as the SF generation process, i.e. $(\omega - n\Delta\omega) + \omega \xrightarrow{}  2\omega - n\Delta\omega$, and serves as an additional loss channel for the mode $\omega - n\Delta\omega$. It is worth mentioning that the weak peaks located at the frequency $\omega - n\Delta\omega/2$ in between the pump and first-order lower sidebands are observed in the primary comb spectra. They are artifacts of the optical spectrum analyzer that records the second-order diffraction of the SF component $2\omega - n\Delta\omega$. 

Moreover, we characterize the $\chi^{(2)}$ grating in the microresonator at each pump detuning denoted in Figs. \ref{Figure3}a and b. The grating images measured at the conditions of SH and 5 FSR comb generation are respectively shown in Figs. \ref{Figure2}a and d, while the other grating images are displayed in Supplementary Note 4. It can be seen that initially there is solely one grating (FH-SH grating), starting from SH generation up to 2 FSR comb. Therefore, the SF coupled primary comb generation process is supported by the same grating as for SH generation. The SH mode involved in the process is identified to be the fundamental TE mode (SH1), whose FSR difference with FH mode is the smallest among SH modes (see Supplementary Note 1). Given the distinct red-shift speeds of FH and SH resonances, such a small FSR difference allows for maximum comb FSR and SF mode switching ($n=1,...,5$) within the pump thermal triangle. 
The secondary grating (FH-SH-TH grating) appears from 3 FSR to 5 FSR cases supporting the cascaded TH generation. The relative amplitude between the two gratings varies with pump detuning, as shown in Supplementary Note 4. 

Figures \ref{Figure3}c and d show the transmission and spectrum measurements for the $1547.35~{\rm nm}$ pump resonance, for which we had recognized the SH mode involved in the SH and SF generation process as the second TE mode (SH2). As a consequence, the FSR difference between the SH and FH is larger compared to the previous $1544.40~{\rm nm}$ resonance (see Supplementary Note 1), and less number of FSR switching events is expected to occur within the pump thermal triangle. Experimentally, we indeed obtain successive primary combs ($n=1,2,3$) with comb spacing up to 3 FSR, and SF component 3 FSR offseted from the SH mode. Apart from the two resonances presented in Fig. \ref{Figure3}, the FH-SH $\chi^{(2)}$ grating inscription and subsequent SF coupled primary comb are found ubiquitous in many pumped resonances within C-band, owing to the high \textit{Q} and small FSR of the microresonator used. 

\begin{figure*}[hbt!]
  \centering{
  \includegraphics[width = 1.0\linewidth]{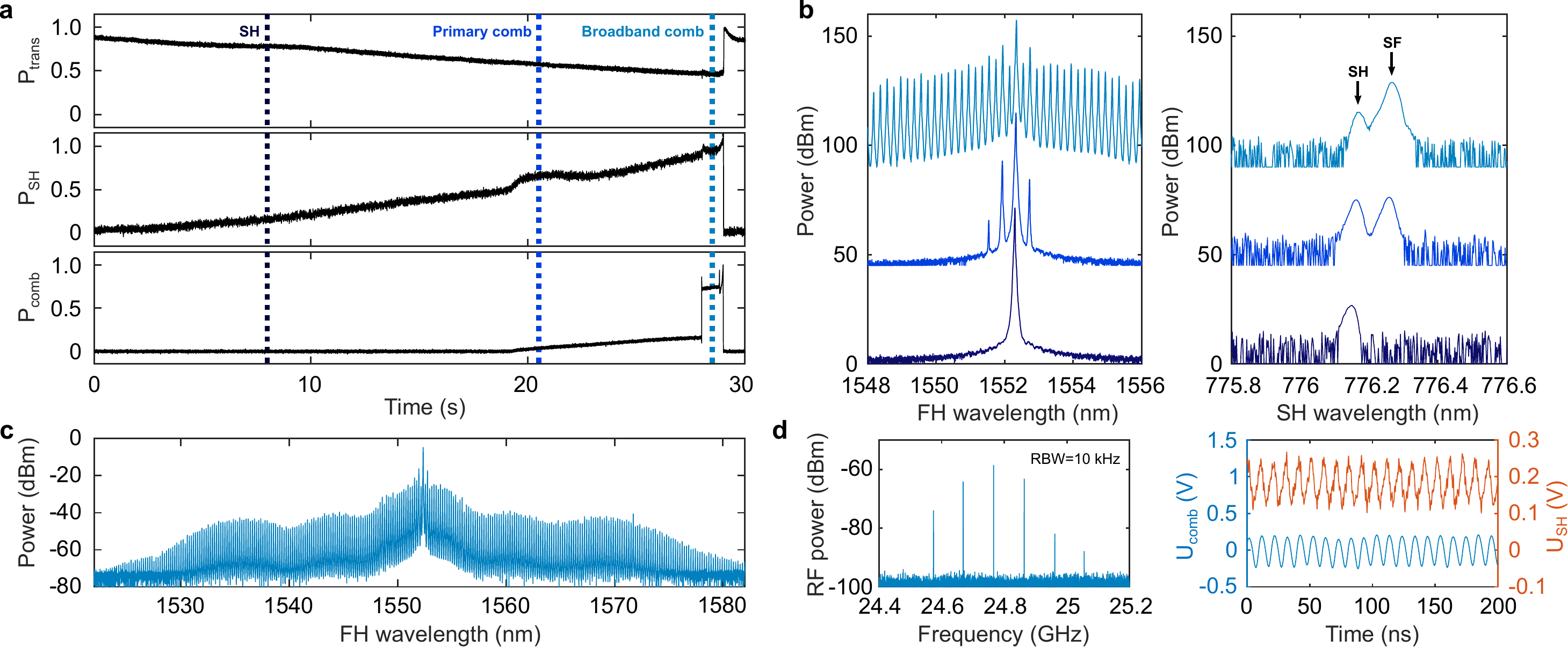}
  } 
    \caption{\noindent\textbf{Generation and characterization of the dark pulse microcomb.} \textbf{a} Evolution of pump transmission ($P_{\rm trans}$), generated SH power ($P_{\rm SH}$) and comb power ($P_{\rm comb}$) during the pump wavelength sweeps across the resonance at $1552.34~{\rm nm}$. \textbf{b} The measured FH (left) and SH (right) spectra at the wavelength regions indicated by the vertical colored lines in \textbf{a}. 
    Initially SH is generated, 2 FSR primary comb is then followed with the simultaneous SF generation, and eventually evolves into a broadband dark pulse state. 
    \textbf{c} The full spectrum of the generated dark pulse state.
    \textbf{d} 
    Repetition-rate beatnote (left) and time dynamics of comb and SH power (right) characterized at the dark pulse state. Multiple peaks spaced by $97.4~{\rm MHz}$ are measured with $10~{\rm kHz}$ resolution bandwidth (RBW), which matches the $\sim 10~{\rm ns}$ oscillation periods of the comb and SH power. The measurements indicate the spectrum in \textbf{c} corresponds to a dark pulse breather state. 
  }
 \label{Figure4}
\end{figure*} 

\vspace{0.1cm}

\noindent\textbf{Broadband microcomb generation.} Besides the aforementioned switching primary combs, broadband microcombs can also be formed in the microresonator with the assistance of SF coupling, as shown in Fig. \ref{Figure4}. Comb generation occurs for the $1552.34~{\rm nm}$ resonance where gratings depicted in Figs. \ref{Figure2}c and f are inscribed. Figure \ref{Figure4}a illustrates the transmission curves of the pump, SH, and comb power when sweeping across the resonance. Three characteristic states are identified in the transmission curves: sole SH generation, 2 FSR primary comb, and the broadband comb state, whose FH and SH spectra are presented in Fig. \ref{Figure4}b from bottom to top, respectively. Once the FH-SH grating is inscribed, it is noted that the 2 FSR primary comb is firstly generated with clear SF coupling at the frequency $\omega - 2\Delta\omega$. We do not observe 1 FSR comb in between the SH generation and 2 FSR comb state. This may be linked to the initial relative resonance detuning condition at FH and SH bands - the nonlinear coupling is not strong enough to give rise to MI for the 1 FSR sideband. Moreover, no FSR switching is cascaded during the pump sweep due to the large FSR mismatch between the FH and the participating SH mode (SH3). Instead, the primary comb evolves directly into a broadband microcomb, whose full spectrum is shown in Fig. \ref{Figure4}c. A typical dark pulse comb shape in the normal dispersion regime is observed, covering the whole C-band with $\sim25~{\rm GHz}$ frequency grid. It is worth mentioning that the dark pulse comb here emerges from the interlocked connection of switching waves\cite{parra2016origin}, rather than the phase-matched $\chi^{(2)}$ process. 
However, the $\chi^{(2)}$ coupling induced MI initiates the formation of primary combs or equivalently the Turing rolls\cite{huang2017globally}. Such temporal patterns indeed facilitate the excitation of the dark pulse states\cite{parra2016origin}, similar to other linear\cite{liu2014investigation,xue2015normal} and nonlinear\cite{xue2017second,cherenkov2017raman} coupling approaches. 

Thanks to the small FSR of the microresonator, we are able to characterize the beatnote of the generated comb using a fast photodetector. 
From the measured RF spectrum in Fig. \ref{Figure4}d we measure multiple peaks equally spaced by $97.4~{\rm MHz}$ with respect to the center frequency $24.77~{\rm GHz}$ (the FSR at FH). The clear beatnote signal indicates that the generated comb is a low-noise breather state\cite{lucas2017breathing,bao2018observation}. Also as shown in Fig. \ref{Figure4}d, we measure the breathing dynamics of the generated comb power (pump filtered) and SH power. Due to the dispersion of the objective lens used, the measurements are done separately by adjusting the distance between the lens and the chip output, therefore their relative oscillation phase is unknown. Nevertheless, it can be seen that the comb and SH power exhibit the same breathing period $\sim 10~{\rm ns}$ well matching the RF spectrum measurement. They are coupled via the SF generation process, where a large breathing depth at SH band is observed (mainly the SF component as indicated by the spectrum in Fig. \ref{Figure4}d). Another interesting feature of the generated dark pulse state is that it can be accessed in a deterministic fashion (see Supplementary Note 5). The same comb state is reached with five consecutive pump sweeps, implying that the current comb excitation route does not undergo the chaotic region\cite{nazemosadat2021switching}.

\vspace{0.1cm}

\section*{Discussion} 
The nonlinear phenomena studied in this work result from the intrinsic $\chi^{(3)}$ and photo-induced $\chi^{(2)}$ nonlinearities in Si$_3$N$_4$ microresonators. Combined $\chi^{(2)}$ and $\chi^{(3)}$ nonlinear effects have been demonstrated in other platforms, for example in aluminum nitride microresonators\cite{jung2014green,bruch2021pockels}, exploiting intermodal phase-matching for the $\chi^{(2)}$ process. In principle these effects should also occur in microresonators with periodic $\chi^{(2)}$ domain inversion\cite{chen2019ultra,lu2020toward}. As opposed to the conventional electric field poling method, the photogalvanic effect recently observed in Si$_3$N$_4$ microresonators has significantly simplified the efforts required to write such QPM structures\cite{nitiss2022optically}. Most importantly, instead of being fixed, the photogalvanic $\chi^{(2)}$ gratings can be reconfigured at different pump resonances, thus offering much more possibility for exploiting new $\chi^{(2)}$ and $\chi^{(3)}$ nonlinear interactions. In this study, we experimentally demonstrate several cascaded nonlinear effects: TH generation via the self-organized FH-SH-TH grating, consecutively switching of primary comb FSR and SF generation within one pump resonance, as well as the broad dark pulse comb generation. 

In particular, the FH-SH-TH grating is appealing as it links three widely separated frequencies at visible, near visible, and telecom bands. 
The simultaneous SH and TH generation has also been previously achieved in $\chi^{(2)}$ crystals\cite{zhang2020dual} and microcavities\cite{zhu1997quasi} by electric field poling, but their underlying domain structures contribute little to the FH-SH-TH grating components. By contrast, the photogalvanic effect self-organizes such a grating with the exactly desired periodicity. In order to fully exploit these distant spectral bands, a specially designed bus waveguide would be needed for efficient SH and TH light coupling. With regards to the FH-SH grating, despite being inscribed by the interference of pump and SH waves (photogalvanic SH generation process), it also automatically provides the phase-matching condition for the SF generation process. This effectively perturbs the phase of the corresponding sideband mode at FH, and serves as an alternative approach for comb initiation in the normal dispersion regime. 
Since the photogalvanic effect in Si$_3$N$_4$ has also been demonstrated at $1~{\rm \mu m}$ band\cite{porcel2017photo}, the method can be adapted towards shorter wavelength range where the dispersion is highly normal dominated by the material dispersion of Si$_3$N$_4$\cite{xue2017second}. 
The simultaneous SH and comb generation is of prominent interest for the self-referencing and the stabilization of the comb. In the current study, only the low-noise dark pulse breather comb is observed but not the stable state, which may be attributed to the high pump power and large detuning used in the experiment condition\cite{bao2018observation,godey2014stability}, and further investigation is needed to better understand the generation of localized states in $\chi^{(3)}$ microresonators with photo-induced $\chi^{(2)}$ nonlinearity.

Owing to the slow time dynamics of the photogalvanic effect\cite{lu2021efficient}, all the pump wavelength sweeps conducted in the experiment are relatively slow, in the time scale of tens of seconds. Therefore, the cascaded nonlinear phenomena presented in this study can also be reliably reproduced by hand tuning the laser frequency. While past works of $\chi^{(2)}$ and $\chi^{(3)}$ combs in Si$_3$N$_4$ microresonators have assumed intermodal phase-matching for the $\chi^{(2)}$ process\cite{miller2014chip,xue2017second}, here we unambiguously confirm the optical inscription of several QPM gratings and observe new cascaded effects of different behaviours depending on the participating SH modes. This work displays the enormous potential of Si$_3$N$_4$ microresonators for the investigation and application of combined $\chi^{(2)}$ and $\chi^{(3)}$ nonlinear effects, in a platform that low-loss and flexibility of optically induced $\chi^{(2)}$ phase-matching can be exploited.

\vspace{0.5cm}

\noindent\textbf{Methods}
\medskip
\begin{footnotesize}

\noindent\textbf{Device information}:
The Si$_3$N$_4$ microresonator we used in the experiment is fabricated by the photonic Damascene process that features ultralow loss operation\cite{liu2021high}. It is a ring structure (radius $R=924~{\rm \mu m}$) coupled with a bus waveguide, both buried in SiO$_2$ cladding. 
The waveguide cross-sections (width $\times$ height) are around $2150~{\rm nm}\times572~{\rm nm}$, which support multiple spatial modes at SH and TH wavelengths (see Supplementary Note 1). At FH wavelength, the intrinsic linewidths and coupling strengths of the fundamental TE mode resonances in the C-band are roughly $30~{\rm MHz}$ and $60~{\rm MHz}$, respectively (see Supplementary Note 1). Besides, the microresonator is characterized to exhibit normal dispersion at pump wavelength. 

\vspace{0.1cm}

\noindent\textbf{TPM imaging and processing}: 
For characterization of the inscribed $\chi^{(2)}$ gratings, a high power femtosecond Ti:Sapphire laser is focused at the grating plane of the microresonator in an upright configuration. The focal spot is then raster-scanned across the plane while, in the meantime, its generated SH signal is monitored so that the $\chi^{(2)}$ response is probed. After TPM images are obtained, we project the circular arc of the microresonator into a straight line so as to facilitate further processing\cite{nitiss2022optically}. Specifically, we set the microresonator's center as the origin and map the point ($\rho \cos\phi$, $\rho \sin\phi$) in the measured image ($\rho$ the distance and $\phi$ the polar angle with respect to the center of the microresonator) to the point ($\rho$, $\rho\phi$) in the new image. Since the grating is written in the microresonator around $\rho = R$, we define $l =\rho\phi$ and $r =\rho - R$ in Fig. \ref{Figure2} for clarity.

\vspace{0.1cm}

\noindent \textbf{Acknowledgements}:
This work was supported by ERC grant PISSARRO (ERC-2017-CoG 771647), by the Air Force Office of Scientific Research (AFOSR) under Award No. FA9550-
19-1-0250, by Contract HR0011-20-2-0046 (NOVEL) from the Defense Advanced Research Projects Agency (DARPA), Microsystems Technology Office (MTO), and by the Swiss National Science Foundation under grant agreement No. 176563 (BRIDGE).

\vspace{0.1cm}

\noindent \textbf{Data Availability Statement}: 
The data and code that support the plots within this paper and other findings of this study are available from the corresponding authors upon reasonable request.
\end{footnotesize}

\bibliographystyle{naturemag}
\bibliography{ref}

\end{document}


\title{Supplementary information for:\\ Photo-induced cascaded harmonic and comb generation in \\ silicon nitride microresonators}

\author{Jianqi Hu}
\thanks{These authors contributed equally to the work.}
\affiliation{{\'E}cole Polytechnique F{\'e}d{\'e}rale de Lausanne, Photonic Systems Laboratory (PHOSL), STI-IEM, Station 11, Lausanne CH-1015, Switzerland.}

\author{Edgars Nitiss}
\thanks{These authors contributed equally to the work.}
\affiliation{{\'E}cole Polytechnique F{\'e}d{\'e}rale de Lausanne, Photonic Systems Laboratory (PHOSL), STI-IEM, Station 11, Lausanne CH-1015, Switzerland.}

\author{Jijun He}
\affiliation{{\'E}cole Polytechnique F{\'e}d{\'e}rale de Lausanne, Laboratory of Photonics and Quantum Measurements (LPQM), SB-IPHYS, Station 3, Lausanne CH-1015, Switzerland.}

\author{Junqiu Liu}
\affiliation{{\'E}cole Polytechnique F{\'e}d{\'e}rale de Lausanne, Laboratory of Photonics and Quantum Measurements (LPQM), SB-IPHYS, Station 3, Lausanne CH-1015, Switzerland.}

\author{Ozan Yakar}
\affiliation{{\'E}cole Polytechnique F{\'e}d{\'e}rale de Lausanne, Photonic Systems Laboratory (PHOSL), STI-IEM, Station 11, Lausanne CH-1015, Switzerland.}

\author{Wenle Weng}
\affiliation{{\'E}cole Polytechnique F{\'e}d{\'e}rale de Lausanne, Laboratory of Photonics and Quantum Measurements (LPQM), SB-IPHYS, Station 3, Lausanne CH-1015, Switzerland.}

\author{Tobias J. Kippenberg}
\affiliation{{\'E}cole Polytechnique F{\'e}d{\'e}rale de Lausanne, Laboratory of Photonics and Quantum Measurements (LPQM), SB-IPHYS, Station 3, Lausanne CH-1015, Switzerland.}

\author{Camille-Sophie Br\`es}
\email[]{camille.bres@epfl.ch}
\affiliation{{\'E}cole Polytechnique F{\'e}d{\'e}rale de Lausanne, Photonic Systems Laboratory (PHOSL), STI-IEM, Station 11, Lausanne CH-1015, Switzerland.}

\maketitle

\section*{\textbf{Supplementary Note 1. Characterization and simulation of the Si$_3$N$_4$ microresonator}}

\begin{figure*}[!b]
  \renewcommand{\figurename}{Supplementary Figure}
  \centering{
  \includegraphics[width = 1.0\linewidth]{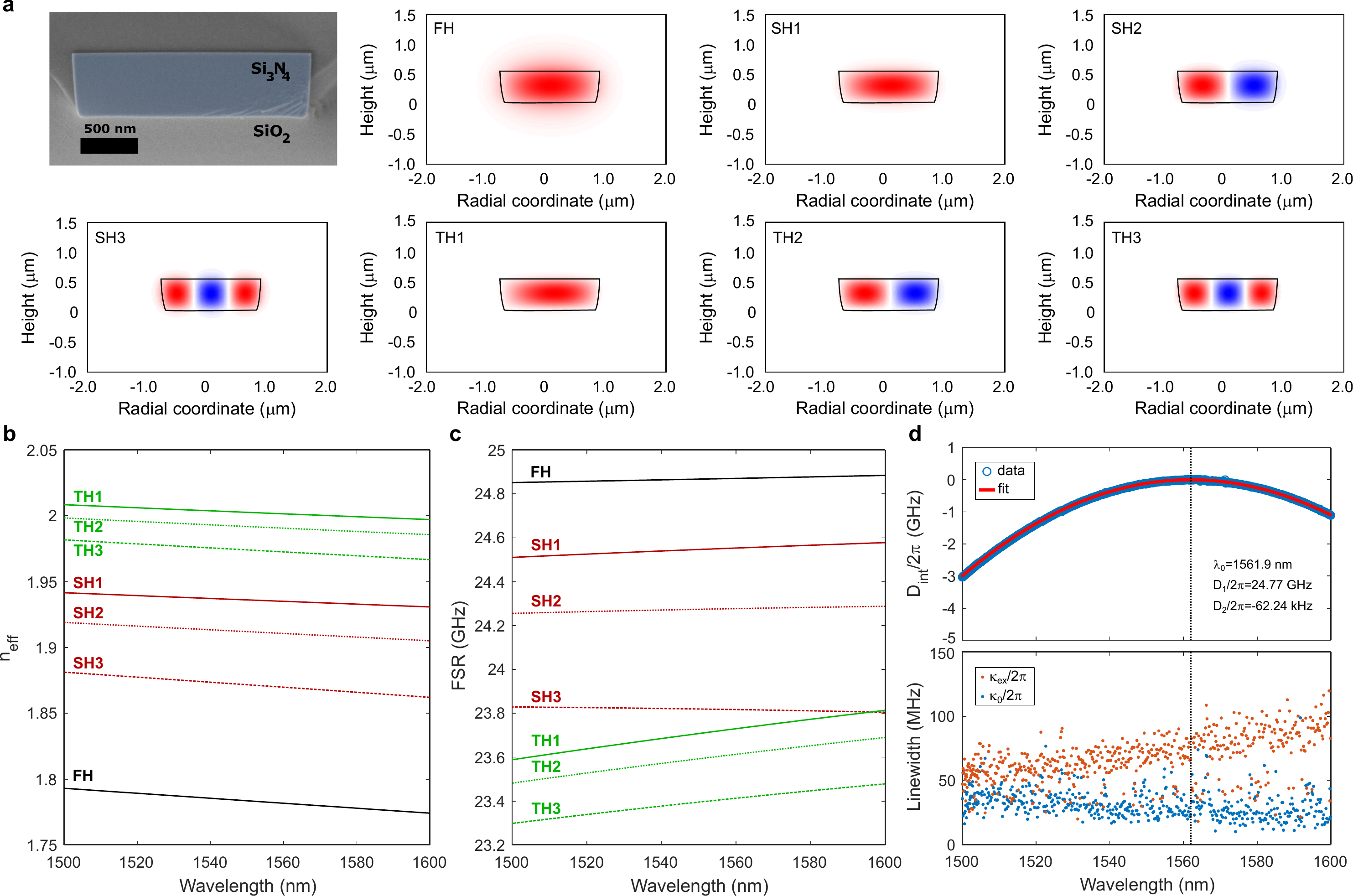}
  } 
    \caption{\noindent\textbf{Characterization and simulation of the Si$_3$N$_4$ microresonator.} \textbf{a} A scanning electron microscope (SEM) image of the waveguide cross-section of the microresonator, along with simulated TE-polarized mode field distributions of fundamental mode at FH wavelength ($\sim1550~{\rm nm}$), first three SH modes (SH1-SH3) and first three TH (TH1-TH3) modes.
    \textbf{b} Simulated effective refractive indices of FH, SH1-SH3, and TH1-TH3 modes with respect to the pump wavelength. 
    \textbf{c} Simulated FSRs of the same modes in \textbf{b} with respect to the pump wavelength. 
    \textbf{d} Top: measured integrated dispersion of the microresonator $D_{\rm int}/2\pi$ and fit. Bottom: measured intrinsic linewidths ($\kappa_{\rm 0}/2\pi$) and coupling strengths ($\kappa_{\rm ex}/2\pi$) of resonances at FH.
  }
 \label{FigureS1}
\end{figure*} 

We utilize a scanning electron microscope (SEM) to accurately retrieve the waveguide cross-section of the used Si$_3$N$_4$ microresonator. 
The measured SEM image is shown in Supplementary Figure \ref{FigureS1}a, where the waveguide core is highlighted, and its actual cross-section (width$\times$height) is estimated to be $2150~{\rm nm}\times572~{\rm nm}$. 
Based on the retrieved waveguide dimension, we use COMSOL Multiphysics to simulate the effective refractive indices of the optical modes involved in this study.
Supplementary Figure \ref{FigureS1}a also displays the simulated TE-polarized mode field distributions for the fundamental mode at FH wavelength ($\sim1550~{\rm nm}$), first three SH modes (SH1 to SH3) as well as first three TH modes (TH1 to TH3). 
For the same set of modes, Supplementary Figures \ref{FigureS1}b and c depict the simulated effective refractive indices and FSR values, respectively, as a function of the pump wavelength in the range from $1500~{\rm nm}$ to $1600~{\rm nm}$. 
The effective refractive indices and field distributions of various modes are essential for $\chi^{(2)}$ grating simulation, while their FSRs are important for the comb study. 
We also employ the frequency-comb-assisted diode laser spectroscopy to measure the resonance linewidths and the dispersion profile of the microresonator\cite{del2009frequency}. 
The top graph in Supplementary Figure \ref{FigureS1}d shows the measured integrated dispersion with respect to the resonance frequency at $\omega_{0}/2\pi=191.94~{\rm THz}$, defined as $D_{\rm int}(\mu)=\omega_{\mu}-\omega_{0}-D_{1} \mu =D_{2}\mu^{2}/2+...~$, where $\omega_{\mu}/2\pi$ is the frequency of the $\mu$-th order resonance, $D_{1}/2\pi=24.77~{\rm GHz}$ is the microresonator's FSR, and $D_{2}/2\pi=-62.24~{\rm kHz}$ is the group-velocity dispersion parameter. 
This indicates the microresonator is normal dispersion in the wavelength range from $1500~{\rm nm}$ to $1600~{\rm nm}$, and no significant avoided mode crossing is observed in the integrated dispersion curve. 
The bottom graph in Supplementary Figure \ref{FigureS1}d shows the measured intrinsic linewidths $\kappa_{\rm 0}/2\pi$ and coupling strengths $\kappa_{\rm ex}/2\pi$ of the microresonator, which are around $30~{\rm MHz}$ and $60~{\rm MHz}$ at the pumped wavelength, respectively.

\section*{\textbf{Supplementary Note 2. Modulational instability gain analysis}}
The FH and SH fields inside the microresonator can be described by mean-field coupled Lugiato–Lefever equations (LLEs)\cite{xue2017second,leo2016frequency}:
\begin{equation}
\frac{\partial A}{\partial t} = -(\frac{\kappa_{\rm a}}{2} + i\delta_{\rm a})A + i \frac{D_{\rm 2,a}}{2}\frac{\partial^2 A}{\partial \phi^2} + i(\gamma_{\rm aa} |A|^2 + \gamma_{\rm ba} |B|^2)A + igBA^{\ast} + \sqrt{\kappa_{\rm ex,a}}s_{\rm in}
\label{lle_a}
\end{equation}
\begin{equation}
\frac{\partial B}{\partial t} = -(\frac{\kappa_{\rm b}}{2} + i\delta_{\rm b})B -\Delta D_1 \frac{\partial B}{\partial \phi} + i \frac{D_{\rm 2,b}}{2}\frac{\partial^2 B}{\partial \phi^2} + i(\gamma_{\rm ab} |A|^2 + \gamma_{\rm bb} |B|^2 )B + ig^{\ast}A^2
\label{lle_b}
\end{equation}
where $A$ and $B$ are respectively the temporal envelopes of intracavity FH and SH in the moving framework of FH, normalized such that their modulus squares correspond to the intacavity photon numbers. $t$ is the slow time axis, $\phi$ is the azimuthal angle along the circumference.
$\sqrt{\kappa_{\rm ex,a}}s_{\rm in}$ represents the external driving term at pump, with $\kappa_{\rm ex,a}$ the FH external coupling rate and $|s_{\rm in}|^2$ the driving power. 
$\kappa_{\rm a(b)}$ and $D_{\rm 2,a(b)}$ denote the total loss rate and dispersion at FH (SH), respectively. $\delta_{\rm a} = \omega_{\rm a} -\omega$ and $\delta_{\rm b} =\omega_{\rm b} -2\omega$ are the detunings at FH and SH, with $\omega_{\rm a,b}$ the cold-cavity resonance frequency closest to $\omega (2\omega)$ and $\omega$ the pump driving frequency. 
$\Delta D_1 = D_{\rm 1,b}-D_{\rm 1,a}$ describes the walk-off between FH and SH.
$\gamma_{\rm aa(bb)}$ and $\gamma_{\rm ab(ba)}$ correspond to the Kerr frequency shifts per photon due to self-phase modulation and cross-phase modulation at FH (SH). $g$ describes the FH-SH coupling strength through the photo-induced $\chi^{(2)}$ QPM grating. 

Reasonable assumptions are made before carrying out the modulational instability (MI) analysis. The dispersion term at SH is dropped out due to the large first-order walk-off term. We also consider the case in the limit of $|A|^2 \gg |B|^2$ so that the Kerr frequency shifts are dominated by the FH power for both FH and SH. Under these assumptions, the stationary solutions of coupled LLEs write:
\begin{equation}
(\frac{\kappa_{\rm a}}{2}+i\delta_{\rm a}-i\gamma_{\rm aa}|A_0|^2 + \frac{|gA_0|^2}{ \frac{\kappa_{\rm b}}{2}+i\delta_{\rm b}-i\gamma_{\rm ab}|A_0|^2} ) A_0 = \sqrt{\kappa_{\rm ex,a}}s_{\rm in}
\label{flat_a}
\end{equation}
\begin{equation}
B_0 = \frac{ig^{\ast}A_0^2}{ \frac{\kappa_{\rm b}}{2}+i\delta_{\rm b}-i\gamma_{\rm ab}|A_0|^2}
\label{flat_b}
\end{equation}
where $A_0$ and $B_0$ are flat solutions of FH and SH, respectively. For MI analysis, we consider the ansatz form of the FH field as $A= A_0 +a_1 e^{i\mu \phi} + a_2 e^{-i\mu \phi}$, where $a_1$ and $a_2$ serve as weak perturbation to $A_0$ at mode indices $\pm \mu$, respectively. 
By assuming the SH field evolves slowly ($\frac{\partial B}{\partial t} \approx 0$)\cite{leo2016frequency}, the perturbation at SH field can be calculated based on Eq. \eqref{lle_b} to the first-order approximation: 
\begin{equation}
\begin{aligned}
B  \approx B_0 +b_1 e^{i\mu \phi} + b_2 e^{-i\mu \phi} 
 = & B_0 + [\gamma_{\rm ab} B_0 (A_0^{\ast}a_1 +A_0 a_2^{\ast})+2g^{\ast} A_0 a_1] J(\mu) e^{i\mu \phi} \\
 & + [\gamma_{\rm ab} B_0 (A_0^{\ast}a_2 +A_0 a_1^{\ast})+2g^{\ast} A_0 a_2] J(-\mu) e^{-i\mu \phi}
\end{aligned}
\label{ansatz_b}
\end{equation}
where we introduce a walk-off function $J(\mu) = \frac{i}{\frac{\kappa_{\rm b}}{2}+i\delta_{\rm b}-i\gamma_{\rm ab}|A_0|^2 +i\Delta D_1 \mu}$ to simplify the expression of $b_1$ and $b_2$.
Notably, the walk-off function $J(\mu)$ is basically a lineshape function with half-linewidth around $\frac{\kappa_{\rm b}}{2}$\cite{leo2016frequency}. It is maximized for a certain $\mu$ when the imaginary part of the denominator is zero ($\delta_{\rm b}-\gamma_{\rm ab}|A_0|^2 +\Delta D_1 \mu = 0$), and decays when the mode index deviates from $\mu$. Therefore, if the FSR mismatch between the FH and SH is much greater than the SH total linewidth ($|\Delta D_1| \gg \frac{\kappa_{\rm b}}{2}$), there is only one integer mode index $\mu$ makes $J(\mu)$ significant. Although we are unable to directly characterize the linewidths of different SH mode resonances, this assumption should be valid in our condition. For lower-order SH modes, the external coupling rates are minimal\cite{nitiss2022optically}, and therefore the total loss rates are dominated by the intrinsic loss rates. 
Suppose the intrinsic loss rate at SH is higher but in the order of that at FH\cite{xue2017second}, the SH loss rate is lower than the FSR difference $|\Delta D_1|$ (see Supplementary Note 1). While for higher-order SH mode $|\Delta D_1|$ reaches $\sim$ GHz for FH-SH3, so that $|\Delta D_1| \gg \frac{\kappa_{\rm b}}{2}$ should still be valid. Thus the walk-off function $J(\mu)$ is only pronounced for one particular mode index $\mu$, and can be considered zero for all the other integers. Note that the steady state SH field can also be written as $B_0 = g^{\ast}A_0^{2}J(0)$. 
Without loss of generality we can omit the terms containing $J(-\mu)$, and considering the sideband generation at SH band is mainly via the sum-frequency process rather than the four-wave mixing involving far detuned SH, Eq. \eqref{ansatz_b} can be simplified as: 
\begin{equation}
B \approx B_0 +2g^{\ast} A_0 a_1 J(\mu) e^{i\mu \phi} 
\label{reduced_b}
\end{equation}

Substitute Eq. \eqref{reduced_b} into Eq. \eqref{lle_a} and separate the modal perturbations in the following form: 
\begin{equation}
{\begin{bmatrix}
\frac{\partial a_1}{\partial t}\\\frac{\partial a_2^{*}}{\partial t}\\
\end{bmatrix}}= 
{\begin{bmatrix}
-\frac{\kappa_{\rm a}}{2}-i\delta_{\rm a}- i\frac{D_{\rm 2,a}}{2}\mu^2 + 2i(\gamma_{\rm aa}+ J(\mu)|g|^2) |A_0|^2  & i\gamma_{\rm aa}A_0^2 + igB_0\\
-i\gamma_{\rm aa}(A_0^{\ast})^2 - ig^{\ast}B_0^{\ast} & -\frac{\kappa_{\rm a}}{2} + i\delta_{\rm a} + i\frac{D_{\rm 2,a}}{2}\mu^2 - 2i\gamma_{\rm aa}|A_0|^2 
\end{bmatrix}}
{\begin{bmatrix}
a_1\\ a_2^{\ast}\\ 
\end{bmatrix}}
\label{matrix}
\end{equation}
The eigenvalue of the matrix determines whether MI could initiate\cite{godey2014stability}:
\begin{equation}
\Gamma(\mu) =  -\frac{\kappa_{\rm a}}{2} + iJ(\mu)|g A_0|^2 \pm \sqrt{|A_0|^4 |\gamma_{\rm aa} +  |g|^2J(0)|^2 - (\delta_a + \frac{D_{\rm 2,a}}{2}\mu^2 -2\gamma_{\rm aa}|A_0|^2 -J(\mu)|g A_0|^2)^2}
\label{eigenvalue}
\end{equation}
When the real part of the eigenvalue is greater than zero (${\rm Re}\{\Gamma(\mu)>0\}$), 
the small perturbation at mode $\mu$ will be effectively amplified, and eventually leads to sideband and SF generation. The structure of computed MI is consistent with ref.\cite{leo2016frequency} if only $\chi^{(2)}$ nonlinearity is considered with the approximation of single-sided coupling ($J(-\mu) \approx 0$). On the other hand, if FH-SH coupling strength $g$ is set to $0$, the derived MI also reduces to the form of conventional $\chi^{(3)}$ microresonators\cite{godey2014stability}.

We use the following values for the MI simulation, i.e., $\kappa_{\rm a}/2\pi = 90~{\rm MHz}$, $\kappa_{\rm ex,a}/2\pi = 60~{\rm MHz}$, $\kappa_{\rm b}/2\pi = 90~{\rm MHz}$, $\Delta D_1/2\pi = -0.32~{\rm GHz}$, $D_{\rm 2,a}/2\pi = -62.24~{\rm kHz}$, $(\omega_{\rm b}-2\omega_{\rm a}) /2\pi = 0.8~{\rm GHz}$, $\gamma_{\rm aa}/2\pi = 0.53~{\rm Hz}$, $\gamma_{\rm ab}/2\pi = 5.09~{\rm Hz}$, $g/2\pi = 1~{\rm kHz}$ and $|s_{\rm in}|^2 = 0.4~{\rm W}$. 
Here we consider the interacting SH to be the fundamental mode and assume the total linewidths of SH resonances are similar to that of FH resonances. 
We chose the cold-cavity resonance offset such that the SH resonance is initially blue-detuned, while being quickly compensated once the pump is coupled into the FH resonance. 
The self-phase and cross-phase nonlinear strengths are calculated based on the simulated mode overlaps\cite{xue2017second,bruch2021pockels}. 
For the FH-SH coupling rate, its magnitude is estimated from the photo-induced $\chi^{(2)}$ measured in Si$_3$N$_4$ microresonators\cite{nitiss2022optically}. The numerically simulated MI gain is shown in Supplementary Figure \ref{FigureS2}. 
Net gain is clearly observed whose peak position in mode index shifts proportionally with the pump detuning, explaining the successive switching of the primary comb FSR. 
While without the FH-SH coupling ($g=0$), net gain is absent in the normal dispersion regime considering the simulation parameters above.

\begin{figure*}[htp]
  \renewcommand{\figurename}{Supplementary Figure}
  \centering{
  \includegraphics[width = 0.6\linewidth]{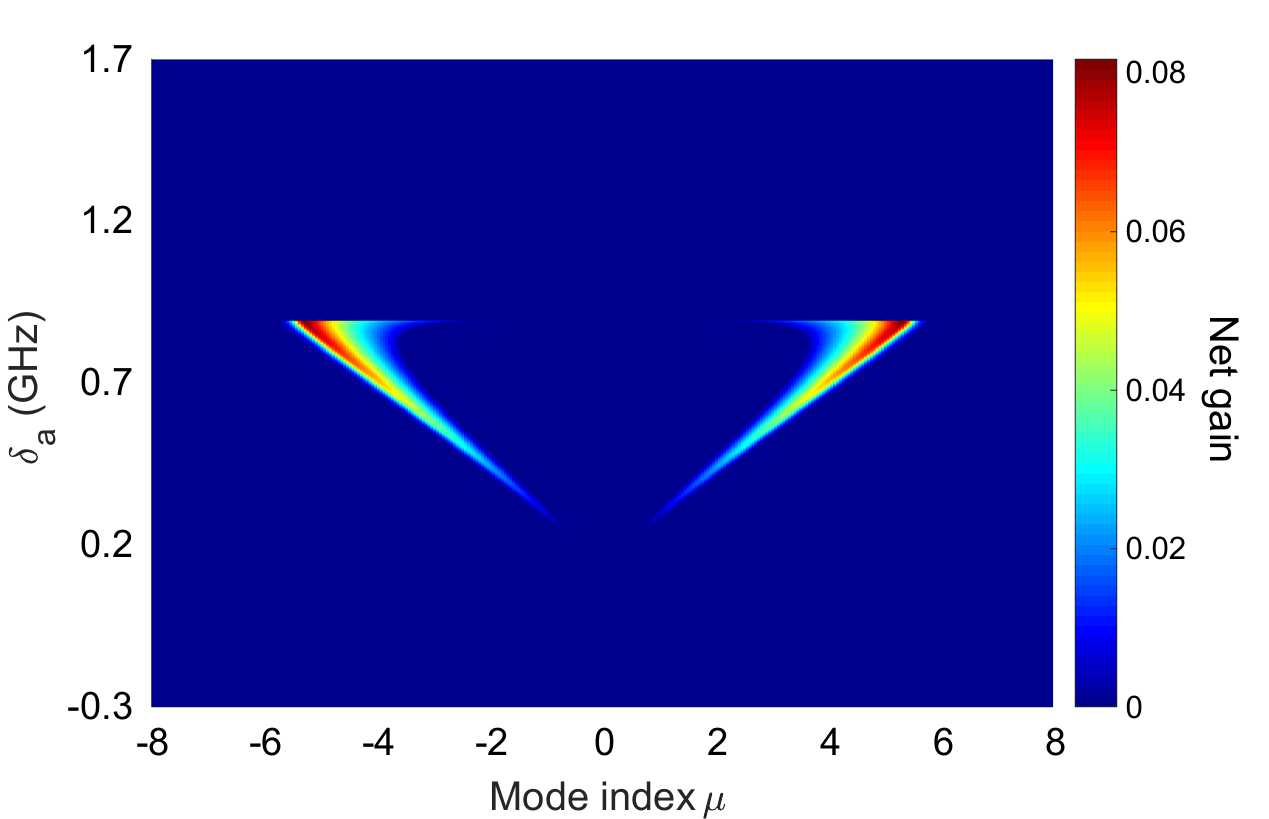}
  } 
    \caption{\noindent\textbf{MI gain analysis for normal dispersion Si$_3$N$_4$ microresonators with FH-SH coupling.} 
    Using the derived MI equation and values present in this Supplementary Note, the net MI gain is simulated as a function of the pump detuning. It can be seen that the peak position of the net gain in mode index shifts proportionally with the pump detuning. 
  }
 \label{FigureS2}
\end{figure*} 

\section*{\textbf{Supplementary Note 3. Experimental setup for comb generation}} 
The experimental setup for optical poling as well as comb generation and characterization in a Si$_3$N$_4$ microresonator is shown in Supplementary Figure \ref{FigureS3}. 
CW light from a C-band tunable laser is amplified and subsequently filtered with a tunable band-pass filter to remove the amplified spontaneous emission.
Then, using a polarization controller and a lensed fiber, the light is coupled to the fundamental TE mode of the bus waveguide. 
The laser frequency is controlled via a function generator, which is slowly decreased so as to stably access the pump resonances of the microresonator.
At the chip output, we collect both the FH and SH light using an optical lens, and separate them via a dichroic beam splitter. The FH light is collimated into an optical fiber and is then split into three parts. 
The first part is directed to a fast photodetector to measure the comb beatnote in an electrical spectrum analyzer. 
For the second part, a Waveshaper is programmed as a notch filter to reject the pump of the FH light, so that both the generated comb power and the pump transmission (the third part) can be measured in an oscilloscope. In terms of SH path, it is split into two parts, which are separately coupled back to optical fibers. 
One half is used to monitor the generated SH power in the oscilloscope, while the other half is directed to an optical spectrum analyzer (OSA). 
Since there is residual FH light in the SH path, the OSA is able to recover both the optical spectra at FH and SH wavelengths. 

\begin{figure*}[htp]
  \renewcommand{\figurename}{Supplementary Figure}
  \centering{
  \includegraphics[width = 0.7\linewidth]{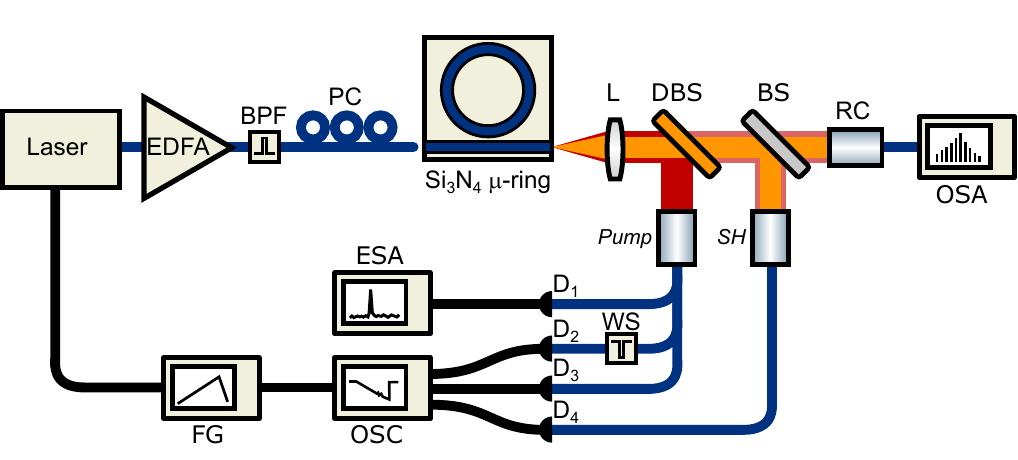}
  } 
    \caption{\noindent\textbf{Experimental setup.} EDFA: erbium-doped fiber amplifier; BPF: band-pass filter; PC: polarization controller; L: lens; DBS: dichroic beam splitter; BS: beam splitter; RC: reflective collimator; D$_1$: high speed FH detector; D$_2$, D$_3$: FH detector; D$_4$: SH detector; OSA: optical spectrum analyzer; FG: function generator; ESA: electrical spectrum analyzer; OSC: oscilloscope; WS: Waveshaper, programmed as a notch filter. Blue lines indicate optical fiber connection, while black lines indicate electrical wiring.
  }
 \label{FigureS3}
\end{figure*} 
 
\section*{\textbf{Supplementary Note 4. Characterization of $\chi^{(2)}$ gratings at different pump detunings}}
\begin{figure*}[!ht]
  \renewcommand{\figurename}{Supplementary Figure}
  \centering{
  \includegraphics[width = 0.9\linewidth]{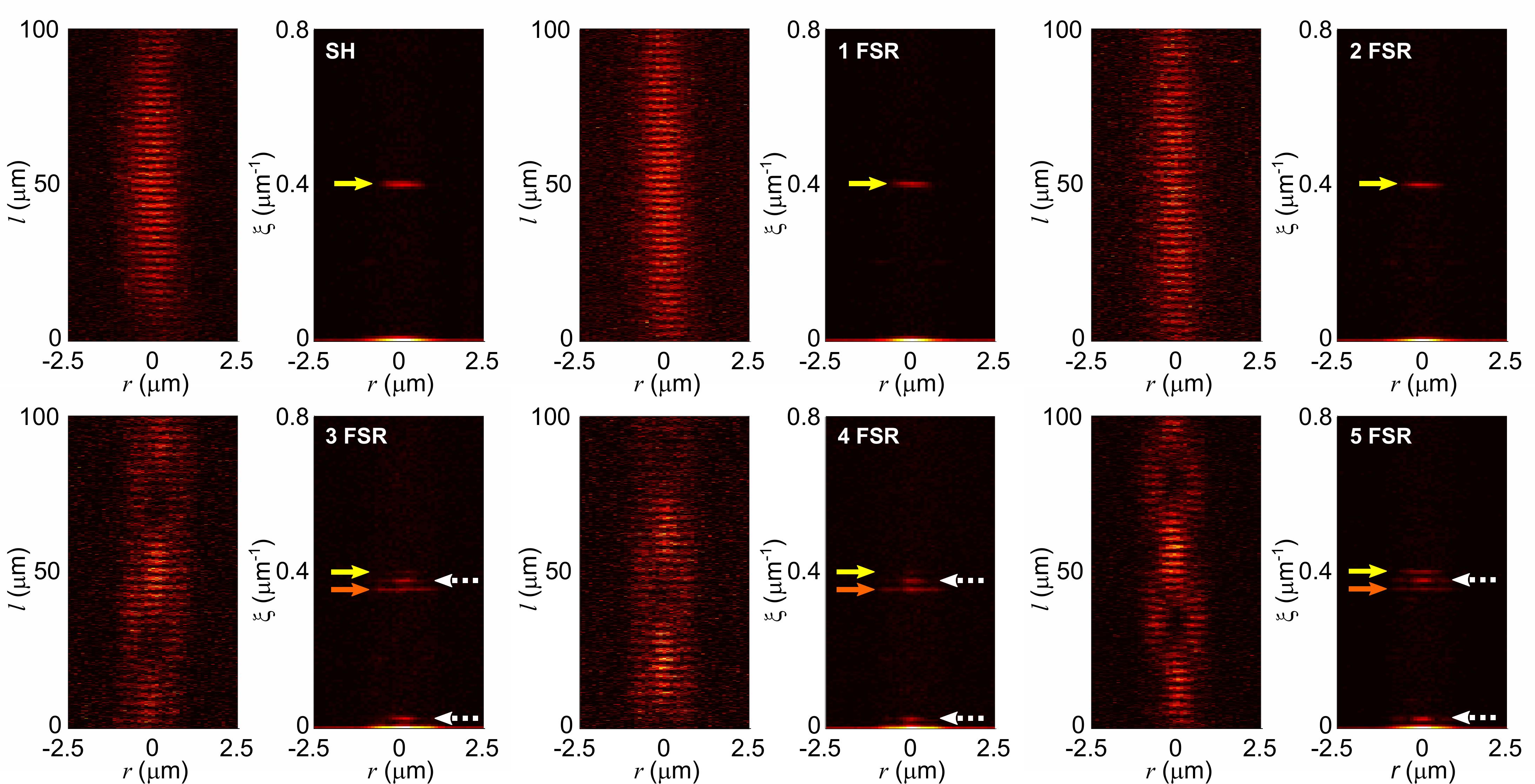}
  } 
    \caption{\noindent\textbf{Grating images and their Fourier analyses at different pump detunings of the 1544.40 nm resonance.} 
    TPM images are measured when SH and SF coupled primary combs of different FSRs (1 FSR to 5 FSR) are generated. 
    The Fourier transform of the images reveals the grating components. For SH, 1 FSR and 2 FSR cases, only FH-SH gratings are inscribed (yellow arrows). 
    While for 3 FSR to 5 FSR cases, in addition to the FH-SH gratings, there are FH-SH-TH gratings (orange arrows) as well as components from the interference of the two gratings (dashed white arrows). Cascaded TH generation is observed for 3 FSR to 5 FSR cases. 
    }
 \label{FigureS4}
\end{figure*} 
In this note, we present the extended TPM characterization of the inscribed $\chi^{(2)}$ gratings for the $1544.40~{\rm nm}$ pump resonance, as shown in Supplementary Figure \ref{FigureS4}. 
Here we measure the $\chi^{(2)}$ gratings at every characteristic detunings from only SH generation to primary combs of different FSRs. 
Note that the grating images in Figs. 2 a and d of the main text correspond to the SH and 5 FSR cases in Supplementary Figure \ref{FigureS4}. At the beginning, a sole FH-SH $\chi^{(2)}$ grating is inscribed in the microresonator enabling SH generation. 
Since the FH-SH grating also phase-matches for the SF generation process, the SF coupled primary comb generation is supported by the same $\chi^{(2)}$ grating.
This is confirmed experimentally by the TPM imaging of the inscribed gratings when the generated SF modes are 1 FSR and 2 FSR downshifted from the SH mode (see Supplementary Figure \ref{FigureS4} 1 FSR and 2 FSR cases). 
For both cases, $\chi^{(2)}$ gratings with the same spatial frequencies (yellow arrows) as the SH case are measured in the microresonator, and no other grating components are observed. 
Further pump detuning leads to the further FSR switching of the SF coupled primary combs as well as the TH generation. 
As shown in the 3 FSR to 5 FSR cases of Supplementary Figure \ref{FigureS4}, we observe superimposed grating structures and analyze them with Fourier transform. 
In addition to the spatial frequency of the initial FH-SH grating, a secondary spatial frequency (orange arrow) and their interference components (dashed white arrows) are identified. 
Here the secondary gratings have the same spatial frequencies for 3 FSR to 5 FSR cases. They are the FH-SH-TH gratings, which phase-match the cascaded TH generation process and are again inscribed via the photogalvanic effect. 

\section*{\textbf{Supplementary Note 5. Deterministic broadband comb generation}}
\begin{figure*}[!ht]
  \renewcommand{\figurename}{Supplementary Figure}
  \centering{
  \includegraphics[width = 0.9\linewidth]{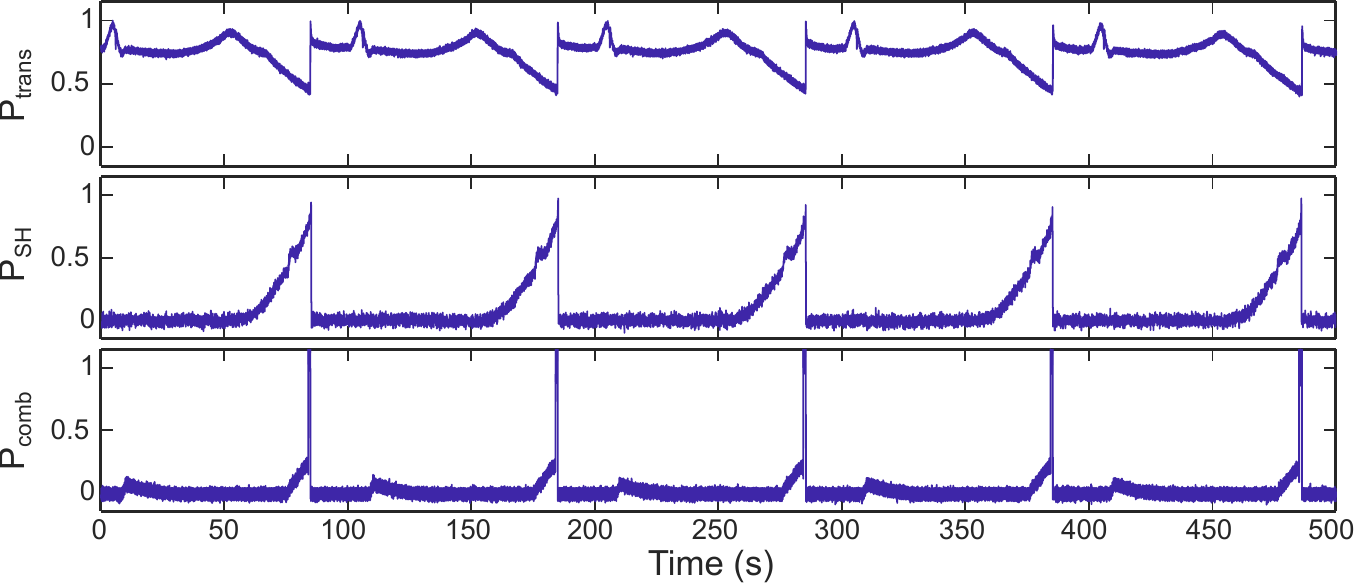}
  } 
    \caption{\noindent\textbf{Deterministic broadband comb generation.} Pump transmission ($P_{\rm trans}$), generated SH power ($P_{\rm SH}$) and comb power ($P_{\rm comb}$) recorded during five consecutive laser frequency scans across the $1552.34~{\rm nm}$ pump resonance. 
    The sharp increase of the measured comb power indicates the generation of the broadband comb state. 
  }
 \label{FigureS5}
\end{figure*}

Here we show that the generated broadband dark pulse microcomb (see Fig. 4 in the main text) can be accessed in a deterministic fashion. Supplementary Figure \ref{FigureS5} illustrates the transmission curves of the pump, SH, and comb power when sweeping across the $1552.34~{\rm nm}$ resonance for five consecutive times. 
In each consecutive sweep, the primary comb at FH is initiated after the SH generation, and is finally evolved into the same broadband comb state in a deterministic manner without transitioning through a chaotic region. 

\bibliographystyle{naturemag}
\bibliography{ref}